\documentclass[conference,10pt,twoside,twocolumn]{IEEEtran}

\usepackage{cite}
\usepackage[T1]{fontenc}
\usepackage{graphicx}
\usepackage{amssymb}
\usepackage{amsmath}
\usepackage{amsthm}
\usepackage{subfigure}
\usepackage{booktabs} 
\usepackage{multirow}
\usepackage{microtype}
\usepackage{balance}
\usepackage{xcolor}
\usepackage{xfrac}    
\usepackage{algorithm}
\usepackage{setspace}
\usepackage{mdframed}
\usepackage{tikz}
\usepackage{circledsteps}

\usepackage{algorithmicx}
\usepackage{algpseudocode}
\algrenewcommand\algorithmicindent{0.7em}%
\usepackage{xpatch}

\usepackage[hidelinks]{hyperref}

\makeatletter
\newcommand\fs@spaceruled{\def\@fs@cfont{\bfseries}\let\@fs@capt\floatc@ruled
  \def\@fs@pre{\vspace{0.1\baselineskip}\hrule height.8pt depth0pt \kern2pt}%
  \def\@fs@post{\kern2pt\hrule\relax}%
  \def\@fs@mid{\kern2pt\hrule\kern2pt}%
  \let\@fs@iftopcapt\iftrue}
\makeatother

\usepackage{amssymb}
\usepackage{amsfonts}
\usepackage{mathrsfs}
\usepackage{xspace}
\usepackage{bm}
\usepackage{upgreek}

\newcommand{\safemath}[2]{\newcommand{#1}{\ensuremath{#2}\xspace}}

\safemath{\bma}{\mathbf{a}}
\safemath{\bmb}{\mathbf{b}}
\safemath{\bmc}{\mathbf{c}}
\safemath{\bmd}{\mathbf{d}}
\safemath{\bme}{\mathbf{e}}
\safemath{\bmf}{\mathbf{f}}
\safemath{\bmg}{\mathbf{g}}
\safemath{\bmh}{\mathbf{h}}
\safemath{\bmi}{\mathbf{i}}
\safemath{\bmj}{\mathbf{j}}
\safemath{\bmk}{\mathbf{k}}
\safemath{\bml}{\mathbf{l}}
\safemath{\bmm}{\mathbf{m}}
\safemath{\bmn}{\mathbf{n}}
\safemath{\bmo}{\mathbf{o}}
\safemath{\bmp}{\mathbf{p}}
\safemath{\bmq}{\mathbf{q}}
\safemath{\bmr}{\mathbf{r}}
\safemath{\bms}{\mathbf{s}}
\safemath{\bmt}{\mathbf{t}}
\safemath{\bmu}{\mathbf{u}}
\safemath{\bmv}{\mathbf{v}}
\safemath{\bmw}{\mathbf{w}}
\safemath{\bmx}{\mathbf{x}}
\safemath{\bmy}{\mathbf{y}}
\safemath{\bmz}{\mathbf{z}}
\safemath{\bmzero}{\mathbf{0}}
\safemath{\bmone}{\mathbf{1}}

\bmdefine{\biad}{a}
\bmdefine{\bibd}{b}
\bmdefine{\bicd}{c}
\bmdefine{\bidd}{d}
\bmdefine{\bied}{e}
\bmdefine{\bifd}{f}
\bmdefine{\bigd}{g}
\bmdefine{\bihd}{h}
\bmdefine{\biid}{i}
\bmdefine{\bijd}{j}
\bmdefine{\bikd}{k}
\bmdefine{\bild}{l}
\bmdefine{\bimd}{m}
\bmdefine{\bind}{n}
\bmdefine{\biod}{o}
\bmdefine{\bipd}{p}
\bmdefine{\biqd}{q}
\bmdefine{\bird}{r}
\bmdefine{\bisd}{s}
\bmdefine{\bitd}{t}
\bmdefine{\biud}{u}
\bmdefine{\bivd}{v}
\bmdefine{\biwd}{w}
\bmdefine{\bixd}{x}
\bmdefine{\biyd}{y}
\bmdefine{\bizd}{z}

\bmdefine{\bixid}{\xi}
\bmdefine{\bilambdad}{\lambda}
\bmdefine{\bimud}{\mu}
\bmdefine{\bithetad}{\theta}
\bmdefine{\biphid}{\phi}
\bmdefine{\bideltad}{\delta}

\safemath{\bmia}{\biad}
\safemath{\bmib}{\bibd}
\safemath{\bmic}{\bicd}
\safemath{\bmid}{\bidd}
\safemath{\bmie}{\bied}
\safemath{\bmif}{\bifd}
\safemath{\bmig}{\bigd}
\safemath{\bmih}{\bihd}
\safemath{\bmii}{\biid}
\safemath{\bmij}{\bijd}
\safemath{\bmik}{\bikd}
\safemath{\bmil}{\bild}
\safemath{\bmim}{\bimd}
\safemath{\bmin}{\bind}
\safemath{\bmio}{\biod}
\safemath{\bmip}{\bipd}
\safemath{\bmiq}{\biqd}
\safemath{\bmir}{\bird}
\safemath{\bmis}{\bisd}
\safemath{\bmit}{\bitd}
\safemath{\bmiu}{\biud}
\safemath{\bmiv}{\bivd}
\safemath{\bmiw}{\biwd}
\safemath{\bmix}{\bixd}
\safemath{\bmiy}{\biyd}
\safemath{\bmiz}{\bizd}

\safemath{\bmxi}{\bixid}
\safemath{\bmlambda}{\bilambdad}
\safemath{\bmmu}{\bimud}
\safemath{\bmtheta}{\bithetad}
\safemath{\bmphi}{\biphid}
\safemath{\bmdelta}{\bideltad}

\safemath{\bA}{\mathbf{A}}
\safemath{\bB}{\mathbf{B}}
\safemath{\bC}{\mathbf{C}}
\safemath{\bD}{\mathbf{D}}
\safemath{\bE}{\mathbf{E}}
\safemath{\bF}{\mathbf{F}}
\safemath{\bG}{\mathbf{G}}
\safemath{\bH}{\mathbf{H}}
\safemath{\bI}{\mathbf{I}}
\safemath{\bJ}{\mathbf{J}}
\safemath{\bK}{\mathbf{K}}
\safemath{\bL}{\mathbf{L}}
\safemath{\bM}{\mathbf{M}}
\safemath{\bN}{\mathbf{N}}
\safemath{\bO}{\mathbf{O}}
\safemath{\bP}{\mathbf{P}}
\safemath{\bQ}{\mathbf{Q}}
\safemath{\bR}{\mathbf{R}}
\safemath{\bS}{\mathbf{S}}
\safemath{\bT}{\mathbf{T}}
\safemath{\bU}{\mathbf{U}}
\safemath{\bV}{\mathbf{V}}
\safemath{\bW}{\mathbf{W}}
\safemath{\bX}{\mathbf{X}}
\safemath{\bY}{\mathbf{Y}}
\safemath{\bZ}{\mathbf{Z}}

\safemath{\bZero}{\mathbf{0}}
\safemath{\bOne}{\mathbf{1}}
\safemath{\bDelta}{\mathbf{\Delta}}
\safemath{\bLambda}{\mathbf{\UpLambda}}
\safemath{\bPhi}{\mathbf{\Upphi}}
\safemath{\bSigma}{\mathbf{\Upsigma}}
\safemath{\bOmega}{\mathbf{\Upomega}}
\safemath{\bTheta}{\mathbf{\Uptheta}}

\bmdefine{\biAd}{A}
\bmdefine{\biBd}{B}
\bmdefine{\biCd}{C}
\bmdefine{\biDd}{D}
\bmdefine{\biEd}{E}
\bmdefine{\biFd}{F}
\bmdefine{\biGd}{G}
\bmdefine{\biHd}{H}
\bmdefine{\biId}{I}
\bmdefine{\biJd}{J}
\bmdefine{\biKd}{K}
\bmdefine{\biLd}{L}
\bmdefine{\biMd}{M}
\bmdefine{\biOd}{N}
\bmdefine{\biPd}{O}
\bmdefine{\biQd}{P}
\bmdefine{\biRd}{R}
\bmdefine{\biSd}{S}
\bmdefine{\biTd}{T}
\bmdefine{\biUd}{U}
\bmdefine{\biVd}{V}
\bmdefine{\biWd}{W}
\bmdefine{\biXd}{X}
\bmdefine{\biYd}{Y}
\bmdefine{\biZd}{Z}

\bmdefine{\biDelta}{\Delta}
\bmdefine{\biLambda}{\Lambda}
\bmdefine{\biPhi}{\Phi}
\bmdefine{\biSigma}{\Sigma}
\bmdefine{\biOmega}{\Omega}
\bmdefine{\biTheta}{\Theta}

\safemath{\bimA}{\biAd}
\safemath{\bimB}{\biBd}
\safemath{\bimC}{\biCd}
\safemath{\bimD}{\biDd}
\safemath{\bimE}{\biEd}
\safemath{\bimF}{\biFd}
\safemath{\bimG}{\biGd}
\safemath{\bimH}{\biHd}
\safemath{\bimI}{\biId}
\safemath{\bimJ}{\biJd}
\safemath{\bimK}{\biKd}
\safemath{\bimL}{\biLd}
\safemath{\bimM}{\biMd}
\safemath{\bimN}{\biNd}
\safemath{\bimO}{\biOd}
\safemath{\bimP}{\biPd}
\safemath{\bimQ}{\biQd}
\safemath{\bimR}{\biRd}
\safemath{\bimS}{\biSd}
\safemath{\bimT}{\biTd}
\safemath{\bimU}{\biUd}
\safemath{\bimV}{\biVd}
\safemath{\bimW}{\biWd}
\safemath{\bimX}{\biXd}
\safemath{\bimY}{\biYd}
\safemath{\bimZ}{\biZd}

\safemath{\bimDelta}{\biDelta}
\safemath{\bimLambda}{\biLambda}
\safemath{\bimPhi}{\biPhi}
\safemath{\bimSigma}{\biSigma}
\safemath{\bimOmega}{\biOmega}
\safemath{\bimTheta}{\biTheta}

\safemath{\setA}{\mathcal{A}}
\safemath{\setB}{\mathcal{B}}
\safemath{\setC}{\mathcal{C}}
\safemath{\setD}{\mathcal{D}}
\safemath{\setE}{\mathcal{E}}
\safemath{\setF}{\mathcal{F}}
\safemath{\setG}{\mathcal{G}}
\safemath{\setH}{\mathcal{H}}
\safemath{\setI}{\mathcal{I}}
\safemath{\setJ}{\mathcal{J}}
\safemath{\setK}{\mathcal{K}}
\safemath{\setL}{\mathcal{L}}
\safemath{\setM}{\mathcal{M}}
\safemath{\setN}{\mathcal{N}}
\safemath{\setO}{\mathcal{O}}
\safemath{\setP}{\mathcal{P}}
\safemath{\setQ}{\mathcal{Q}}
\safemath{\setR}{\mathcal{R}}
\safemath{\setS}{\mathcal{S}}
\safemath{\setT}{\mathcal{T}}
\safemath{\setU}{\mathcal{U}}
\safemath{\setV}{\mathcal{V}}
\safemath{\setW}{\mathcal{W}}
\safemath{\setX}{\mathcal{X}}
\safemath{\setY}{\mathcal{Y}}
\safemath{\setZ}{\mathcal{Z}}
\safemath{\emptySet}{\varnothing}

\safemath{\colA}{\mathscr{A}}
\safemath{\colB}{\mathscr{B}}
\safemath{\colC}{\mathscr{C}}
\safemath{\colD}{\mathscr{D}}
\safemath{\colE}{\mathscr{E}}
\safemath{\colF}{\mathscr{F}}
\safemath{\colG}{\mathscr{G}}
\safemath{\colH}{\mathscr{H}}
\safemath{\colI}{\mathscr{I}}
\safemath{\colJ}{\mathscr{J}}
\safemath{\colK}{\mathscr{K}}
\safemath{\colL}{\mathscr{L}}
\safemath{\colM}{\mathscr{M}}
\safemath{\colN}{\mathscr{N}}
\safemath{\colO}{\mathscr{O}}
\safemath{\colP}{\mathscr{P}}
\safemath{\colQ}{\mathscr{Q}}
\safemath{\colR}{\mathscr{R}}
\safemath{\colS}{\mathscr{S}}
\safemath{\colT}{\mathscr{T}}
\safemath{\colU}{\mathscr{U}}
\safemath{\colV}{\mathscr{V}}
\safemath{\colW}{\mathscr{W}}
\safemath{\colX}{\mathscr{X}}
\safemath{\colY}{\mathscr{Y}}
\safemath{\colZ}{\mathscr{Z}}

\safemath{\opA}{\mathbb{A}}
\safemath{\opB}{\mathbb{B}}
\safemath{\opC}{\mathbb{C}}
\safemath{\opD}{\mathbb{D}}
\safemath{\opE}{\mathbb{E}}
\safemath{\opF}{\mathbb{F}}
\safemath{\opG}{\mathbb{G}}
\safemath{\opH}{\mathbb{H}}
\safemath{\opI}{\mathbb{I}}
\safemath{\opJ}{\mathbb{J}}
\safemath{\opK}{\mathbb{K}}
\safemath{\opL}{\mathbb{L}}
\safemath{\opM}{\mathbb{M}}
\safemath{\opN}{\mathbb{N}}
\safemath{\opO}{\mathbb{O}}
\safemath{\opP}{\mathbb{P}}
\safemath{\opQ}{\mathbb{Q}}
\safemath{\opR}{\mathbb{R}}
\safemath{\opS}{\mathbb{S}}
\safemath{\opT}{\mathbb{T}}
\safemath{\opU}{\mathbb{U}}
\safemath{\opV}{\mathbb{V}}
\safemath{\opW}{\mathbb{W}}
\safemath{\opX}{\mathbb{X}}
\safemath{\opY}{\mathbb{Y}}
\safemath{\opZ}{\mathbb{Z}}
\safemath{\opZero}{\mathbb{O}}
\safemath{\identityop}{\opI}

\safemath{\veca}{\bma}
\safemath{\vecb}{\bmb}
\safemath{\vecc}{\bmc}
\safemath{\vecd}{\bmd}
\safemath{\vece}{\bme}
\safemath{\vecf}{\bmf}
\safemath{\vecg}{\bmg}
\safemath{\vech}{\bmh}
\safemath{\veci}{\bmi}
\safemath{\vecj}{\bmj}
\safemath{\veck}{\bmk}
\safemath{\vecl}{\bml}
\safemath{\vecm}{\bmm}
\safemath{\vecn}{\bmn}
\safemath{\veco}{\bmo}
\safemath{\vecp}{\bmp}
\safemath{\vecq}{\bmq}
\safemath{\vecr}{\bmr}
\safemath{\vecs}{\bms}
\safemath{\vect}{\bmt}
\safemath{\vecu}{\bmu}
\safemath{\vecv}{\bmv}
\safemath{\vecw}{\bmw}
\safemath{\vecx}{\bmx}
\safemath{\vecy}{\bmy}
\safemath{\vecz}{\bmz}

\safemath{\veczero}{\bmzero}
\safemath{\vecone}{\bmone}
\safemath{\vecxi}{\bmxi}
\safemath{\veclambda}{\bmlambda}
\safemath{\vecmu}{\bmmu}
\safemath{\vectheta}{\bmtheta}
\safemath{\vecphi}{\bmphi}
\safemath{\vecdelta}{\bmdelta}

\safemath{\matA}{\bA}
\safemath{\matB}{\bB}
\safemath{\matC}{\bC}
\safemath{\matD}{\bD}
\safemath{\matE}{\bE}
\safemath{\matF}{\bF}
\safemath{\matG}{\bG}
\safemath{\matH}{\bH}
\safemath{\matI}{\bI}
\safemath{\matJ}{\bJ}
\safemath{\matK}{\bK}
\safemath{\matL}{\bL}
\safemath{\matM}{\bM}
\safemath{\matN}{\bN}
\safemath{\matO}{\bO}
\safemath{\matP}{\bP}
\safemath{\matQ}{\bQ}
\safemath{\matR}{\bR}
\safemath{\matS}{\bS}
\safemath{\matT}{\bT}
\safemath{\matU}{\bU}
\safemath{\matV}{\bV}
\safemath{\matW}{\bW}
\safemath{\matX}{\bX}
\safemath{\matY}{\bY}
\safemath{\matZ}{\bZ}
\safemath{\matzero}{\bmzero}

\safemath{\matDelta}{\bDelta}
\safemath{\matLambda}{\bLambda}
\safemath{\matPhi}{\bPhi}
\safemath{\matSigma}{\bSigma}
\safemath{\matOmega}{\bOmega}
\safemath{\matTheta}{\bTheta}

\safemath{\matidentity}{\matI}
\safemath{\matone}{\matO}

\safemath{\rnda}{A}
\safemath{\rndb}{B}
\safemath{\rndc}{C}
\safemath{\rndd}{D}
\safemath{\rnde}{E}
\safemath{\rndf}{F}
\safemath{\rndg}{G}
\safemath{\rndh}{H}
\safemath{\rndi}{I}
\safemath{\rndj}{J}
\safemath{\rndk}{K}
\safemath{\rndl}{L}
\safemath{\rndm}{M}
\safemath{\rndn}{N}
\safemath{\rndo}{O}
\safemath{\rndp}{P}
\safemath{\rndq}{Q}
\safemath{\rndr}{R}
\safemath{\rnds}{S}
\safemath{\rndt}{T}
\safemath{\rndu}{U}
\safemath{\rndv}{V}
\safemath{\rndw}{W}
\safemath{\rndx}{X}
\safemath{\rndy}{Y}
\safemath{\rndz}{Z}

\safemath{\rveca}{\bimA}
\safemath{\rvecb}{\bimB}
\safemath{\rvecc}{\bimC}
\safemath{\rvecd}{\bimD}
\safemath{\rvece}{\bimE}
\safemath{\rvecf}{\bimF}
\safemath{\rvecg}{\bimG}
\safemath{\rvech}{\bimH}
\safemath{\rveci}{\bimI}
\safemath{\rvecj}{\bimJ}
\safemath{\rveck}{\bimK}
\safemath{\rvecl}{\bimL}
\safemath{\rvecm}{\bimM}
\safemath{\rvecn}{\bimN}
\safemath{\rveco}{\bomO}
\safemath{\rvecp}{\bimP}
\safemath{\rvecq}{\bimQ}
\safemath{\rvecr}{\bimR}
\safemath{\rvecs}{\bimS}
\safemath{\rvect}{\bimT}
\safemath{\rvecu}{\bimU}
\safemath{\rvecv}{\bimV}
\safemath{\rvecw}{\bimW}
\safemath{\rvecx}{\bimX}
\safemath{\rvecy}{\bimY}
\safemath{\rvecz}{\bimZ}

\safemath{\rvecxi}{\bmxi}
\safemath{\rveclambda}{\bmlambda}
\safemath{\rvecmu}{\bmmu}
\safemath{\rvectheta}{\bmtheta}
\safemath{\rvecphi}{\bmphi}

\safemath{\rmatA}{\bimA}
\safemath{\rmatB}{\bimB}
\safemath{\rmatC}{\bimC}
\safemath{\rmatD}{\bimD}
\safemath{\rmatE}{\bimE}
\safemath{\rmatF}{\bimF}
\safemath{\rmatG}{\bimG}
\safemath{\rmatH}{\bimH}
\safemath{\rmatI}{\bimI}
\safemath{\rmatJ}{\bimJ}
\safemath{\rmatK}{\bimK}
\safemath{\rmatL}{\bimL}
\safemath{\rmatM}{\bimM}
\safemath{\rmatN}{\bimN}
\safemath{\rmatO}{\bimO}
\safemath{\rmatP}{\bimP}
\safemath{\rmatQ}{\bimQ}
\safemath{\rmatR}{\bimR}
\safemath{\rmatS}{\bimS}
\safemath{\rmatT}{\bimT}
\safemath{\rmatU}{\bimU}
\safemath{\rmatV}{\bimV}
\safemath{\rmatW}{\bimW}
\safemath{\rmatX}{\bimX}
\safemath{\rmatY}{\bimY}
\safemath{\rmatZ}{\bimZ}

\safemath{\rmatDelta}{\bimDelta}
\safemath{\rmatLambda}{\bimLambda}
\safemath{\rmatPhi}{\bimPhi}
\safemath{\rmatSigma}{\bimSigma}
\safemath{\rmatOmega}{\bimOmega}
\safemath{\rmatTheta}{\bimTheta}

\usepackage{amssymb}
\usepackage{amsfonts}
\usepackage{mathrsfs}
\usepackage{xspace}
\usepackage{bm}
\usepackage{fancyref}
\usepackage{textcomp}

\usepackage{multirow}
\usepackage{stmaryrd}

\newenvironment{textbmatrix}{	\setlength{\arraycolsep}{2.5pt}%
								\big[\begin{matrix}}{\end{matrix}\big]%
								\raisebox{0.08ex}{\vphantom{M}}}

\def\be{\begin{equation}}
\def\ee{\end{equation}}
\def\een{\nonumber \end{equation}}
\def\mat{\begin{bmatrix}}
\def\emat{\end{bmatrix}}
\def\btm{\begin{textbmatrix}}
\def\etm{\end{textbmatrix}}

\def\ba#1\ea{\begin{align}#1\end{align}}
\def\bas#1\eas{\begin{align*}#1\end{align*}}
\def\bs#1\es{\begin{split}#1\end{split}}
\def\bg#1\eg{\begin{gather}#1\end{gather}}
\def\bml#1\eml{\begin{multline}#1\end{multline}}
\def\bi#1\ei{\begin{itemize}#1\end{itemize}}

\newcommand{\lefto}{\mathopen{}\left}

\DeclareMathOperator{\Exop}{\opE}			

\newcommand{\orth}{\perp}					
\newcommand{\Ex}[2]{\ensuremath{\Exop_{#1}\lefto[#2\right]}} 	

\newcommand{\herm}[1]{\ensuremath{#1^{\text{H}}}} 	
\newcommand{\pinv}[1]{\ensuremath{#1^{\dagger}}} 	

\safemath{\dirac}{\delta}					
\safemath{\krond}{\dirac}					

\safemath{\upto}{\uparrow}
\safemath{\downto}{\downarrow}
\safemath{\iu}{j}							
\safemath{\ev}{\lambda}						
\safemath{\hilseqspace}{l^{2}}				
\newcommand{\banachfunspace}[1]{\setL^{#1}}	
\safemath{\hilfunspace}{\banachfunspace{2}}	

\safemath{\SNR}{\textit{SNR}} 				
\safemath{\PAR}{\textit{PAR}} 				
\safemath{\No}{N_0}							
\safemath{\Es}{E_s}							
\safemath{\Eb}{E_b}							
\safemath{\EbNo}{\frac{\Eb}{\No}}
\safemath{\EsNo}{\frac{\Es}{\No}}

\DeclareMathOperator{\CHop}{\ensuremath{\opH}} 
\safemath{\tvir}{\rndh_{\CHop}}				
\safemath{\tvtf}{\rndl_{\CHop}}				
\safemath{\spf}{\rnds_{\CHop}}				
\safemath{\bff}{H_{\CHop}}					

\safemath{\ircf}{r_{h}}						
\safemath{\tftvcf}{r_{s}}					
\safemath{\tfcf}{r_{l}}						
\safemath{\bfcf}{r_{H}}						

\safemath{\tcorr}{c_h}						
\safemath{\scf}{c_{s}}						
\safemath{\tfcorr}{c_{l}}					
\safemath{\fcorr}{c_{H}}						

\safemath{\mi}{I}							
\safemath{\capacity}{C}						

\safemath{\normal}{\mathcal{N}}			
\safemath{\jpg}{\mathcal{CN}}			
\safemath{\mchain}{\leftrightarrow}		

\safemath{\dB}{\,\mathrm{dB}}
\safemath{\dBm}{\,\mathrm{dBm}}
\safemath{\Hz}{\,\mathrm{Hz}}
\safemath{\kHz}{\,\mathrm{kHz}}
\safemath{\MHz}{\,\mathrm{MHz}}
\safemath{\GHz}{\,\mathrm{GHz}}
\safemath{\s}{\,\mathrm{s}}
\safemath{\ms}{\,\mathrm{ms}}
\safemath{\mus}{\,\mathrm{\text{\textmu}s}}
\safemath{\ns}{\,\mathrm{ns}}
\safemath{\ps}{\,\mathrm{ps}}
\safemath{\meter}{\,\mathrm{m}}
\safemath{\mm}{\,\mathrm{mm}}
\safemath{\cm}{\,\mathrm{cm}}
\safemath{\m}{\,\mathrm{m}}
\safemath{\W}{\,\mathrm{W}}
\safemath{\mW}{\, \mathrm{mW}}
\safemath{\J}{\,\mathrm{J}}
\safemath{\K}{\,\mathrm{K}}
\safemath{\bit}{\,\mathrm{bit}}
\safemath{\nat}{\,\mathrm{nat}}

\safemath{\define}{\triangleq}			

\safemath{\equivalent}{\sim}
\safemath{\distas}{\sim}					
\safemath{\sdiff}{\Delta}				

\safemath{\reals}{\mathbb{R}}
\safemath{\positivereals}{\reals_{+}}
\safemath{\integers}{\mathbb{Z}}
\safemath{\posint}{\integers_{+}}
\safemath{\naturals}{\mathbb{N}}
\safemath{\posnaturals}{\naturals_{+}}
\safemath{\complexset}{\mathbb{C}}
\safemath{\rationals}{\mathbb{Q}}

\newcommand*{\fancyrefapplabelprefix}{app}		
\newcommand*{\fancyrefthmlabelprefix}{thm}		
\newcommand*{\fancyreflemlabelprefix}{lem}		
\newcommand*{\fancyrefcorlabelprefix}{cor}		
\newcommand*{\fancyrefdeflabelprefix}{def}		
\newcommand*{\fancyrefproplabelprefix}{prop}		
\newcommand*{\fancyrefexmpllabelprefix}{exmpl}
\newcommand*{\fancyrefalglabelprefix}{alg}		
\newcommand*{\fancyreftbllabelprefix}{tbl}		

\frefformat{vario}{\fancyrefseclabelprefix}{Sec.~#1}
\frefformat{vario}{\fancyrefthmlabelprefix}{Thm.~#1}
\frefformat{vario}{\fancyreftbllabelprefix}{Table~#1}
\frefformat{vario}{\fancyreflemlabelprefix}{Lemma~#1}
\frefformat{vario}{\fancyrefcorlabelprefix}{Corollary~#1}
\frefformat{vario}{\fancyrefdeflabelprefix}{Def.~#1}
\frefformat{vario}{\fancyreffiglabelprefix}{Fig.~#1}
\frefformat{vario}{\fancyrefapplabelprefix}{Appendix~#1}
\frefformat{vario}{\fancyrefeqlabelprefix}{(#1)}
\frefformat{vario}{\fancyrefproplabelprefix}{Prop.~#1}
\frefformat{vario}{\fancyrefexmpllabelprefix}{Example~#1}
\frefformat{vario}{\fancyrefalglabelprefix}{Alg.~#1}

 \newtheorem{thm}{Theorem}

 \newtheorem*{remark*}{Remark}

\safemath{\dictab}{[\,\dicta\,\,\dictb\,]}

\safemath{\ysig}{\bmy}
\safemath{\ysighat}{\hat{\ysig}}
\safemath{\ysigdim}{M}
\safemath{\xsig}{\bmx}
\safemath{\xsigdim}{N}
\safemath{\nx}{n_x}
\safemath{\zsig}{\bmz}
\safemath{\zsigdim}{\ysigdim}
\safemath{\rsig}{\bmr}
\safemath{\Adict}{\bA}
\safemath{\Adicttilde}{\widetilde{\Adict}}
\safemath{\Adictdim}{\outputdim\times\xsigdim}
\safemath{\avec}{\bma}
\safemath{\avectilde}{\tilde{\avec}}
\safemath{\Bdict}{\bB}
\safemath{\Bdicttilde}{\widetilde{\Bdict}}
\safemath{\Cdict}{\bC}
\safemath{\cvec}{\bmc}
\safemath{\Ddict}{\bD}
\safemath{\Ddictdim}{\ysigdim\times\xsigdim}
\safemath{\dvec}{\bmd}
\safemath{\Ddicttilde}{\widetilde{\bD}}
\safemath{\Bonb}{\bB}
\safemath{\bvec}{\bmb}
\safemath{\Bonbdim}{\ysigdim\times\ysigdim}
\safemath{\noise}{\bmn}
\safemath{\noisedim}{\ysigim}
\safemath{\err}{\bme}
\safemath{\errdim}{\ysigdim}
\safemath{\errset}{\setE}
\safemath{\nerr}{n_e}
\safemath{\delop}{\bP_\errset}
\safemath{\delopc}{\bP_{{\errset}^c}}

\safemath{\cplxi}{\imath}
\safemath{\cplxj}{\jmath}

\safemath{\dict}{\matD}
\safemath{\inputdim}{N}		
\safemath{\outputdim}{M}		
\safemath{\sparsity}{S}	
\safemath{\inputdimA}{{N_a}}	
\safemath{\inputdimB}{{N_b}}	
\safemath{\elemA}{{n_a}}	
\safemath{\elemB}{{n_b}}	
\safemath{\resA}{\matR_a}	
\safemath{\resB}{\matR_b}	
\safemath{\subD}{\matS} 
\safemath{\subA}{\matS_a} 
\safemath{\subB}{\matS_b} 
\safemath{\dicta}{\matA} 	
\safemath{\dictb}{\matB} 	
\safemath{\hollowS}{H}
\safemath{\hollowA}{H_a}
\safemath{\hollowB}{H_b}
\safemath{\cross}{Z}
\safemath{\coh}{\mu_d}			
\safemath{\coha}{\mu_a}			
\safemath{\cohb}{\mu_b}			
\safemath{\mubs}{\nu}	
\safemath{\cohm}{\mu_m} 
\safemath{\dictset}{\setD}	
\safemath{\dictsetp}{\dictset(\coh,\coha,\cohb)}	
\safemath{\dictsetgen}{\dictset_\text{gen}}
\safemath{\dictsetgenp}{\dictsetgen(\coh)}
\safemath{\dictsetonb}{\dictset_\text{onb}}
\safemath{\dictsetonbp}{\dictsetonb(\coh)}

\safemath{\leftside}{U}
\safemath{\rightsideA}{R_a}
\safemath{\rightsideB}{R_b}

\safemath{\indexS}{\setI_S} 

\safemath{\na}{n_a}			
\safemath{\nb}{n_b}			
\safemath{\coeffa}{p_i}	
\safemath{\coeffb}{q_j}	
\safemath{\seta}{\setP}		
\safemath{\setb}{\setQ}     
\safemath{\setw}{\setW}	
\safemath{\setz}{\setZ}	
\safemath{\cola}{\veca}		
\safemath{\colb}{\vecb}		
\safemath{\cold}{\vecd}		
\safemath{\inputvec}{\vecx} 	
\safemath{\error}{\vece}	
\safemath{\noiseout}{\vecz} 	
\safemath{\inputvecel}{x}
\safemath{\inputveca}{\vecx_a}
\safemath{\inputvecb}{\vecx_b}
\safemath{\outputvec}{\vecy}	
\safemath{\lambdamin}{\lambda_{\mathrm{min}}}

\safemath{\elltwo}{\ell_2}
\safemath{\ellone}{\ell_1}
\safemath{\ellzero}{\ell_0}
\safemath{\ellinf}{\ell_\infty}
\safemath{\ellinftilde}{\ell_{\widetilde\infty}}
\safemath{\licard}{Z(\coh,\coha,\cohb)}
\safemath{\xsol}{\hat{x}}
\safemath{\xbord}{x_b}		
\safemath{\xstat}{x_s}		
\safemath{\xstatLone}{\tilde{x}_s}
\safemath{\order}{\mathcal{O}} 
\safemath{\scales}{\Theta} 
\safemath{\ones}{\mathbf{1}} 
\safemath{\zeroes}{\mathbf{0}} 
\safemath{\thlone}{\kappa(\coh,\cohb)} 
\safemath{\constoneA}{\delta} 
\safemath{\constoneB}{\epsilon} 
\safemath{\nlarge}{L}				   
\safemath{\sumlarge}{S_\nlarge}
\safemath{\maxlarger}{P_\nlarge}	   
\safemath{\Pzero}{\textrm{P0}}	
\safemath{\Pone}{\textrm{P1}}
\safemath{\vecfir}{\vecw}			 
\safemath{\vecsec}{\vecz}
\safemath{\elvecfir}{w}              
\safemath{\elvecsec}{z}				 
\safemath{\nlargefir}{n}
\safemath{\normout}{\gamma}
\safemath{\auxfun}{h}
\safemath{\supp}{\textrm{supp}}

\safemath{\indexa}{\ell}
\safemath{\indexb}{r}
\safemath{\indexc}{i}
\safemath{\indexd}{j}

\safemath{\project}{P}

\usepackage{framed}

\IEEEoverridecommandlockouts
\allowdisplaybreaks 

\definecolor{gray}{HTML}{555555}   
\newcommand*\graycircled[1]{\Circled[inner color=white, fill color= gray, outer color=gray]{\small{\textnormal{#1}}}}
\newcommand*\tinygraycircled[1]{\Circled[inner color=white, fill color= gray, outer color=gray]{\scriptsize{#1}}}

\safemath{\Hj}{\bJ}
\safemath{\bsj}{\bmw}
\safemath{\sj}{w}
\safemath{\Ej}{E_w}
\safemath{\proxg}{\text{prox}_g}
\safemath{\rE}{\rho_{\textsf{\tiny{E}}}}
\safemath{\rP}{\rho_{\textsf{\tiny{P}}}}

\begin{document}
\bstctlcite{IEEEexample:BSTcontrol} 

\title{Joint Jammer Mitigation and Data Detection for\\
Smart, Distributed, and Multi-Antenna Jammers}

\author{\IEEEauthorblockN{Gian Marti and Christoph Studer}
\IEEEauthorblockA{\em Department of Information Technology
and Electrical Engineering, ETH Zurich, Switzerland\\
email: gimarti@ethz.ch and studer@ethz.ch 
}
\thanks{The work of CS was supported in part by the U.S.\ National Science Foundation (NSF) under grants CNS-1717559 and ECCS-1824379. The work of GM and CS was supported in part by an ETH Research~Grant.}
\thanks{The authors thank Sueda Taner for comments and suggestions.}
}

\maketitle

\begin{abstract}
Multi-antenna (MIMO) processing is a promising solution to the problem of jammer mitigation. Existing methods mitigate the jammer based on an estimate of its subspace (or receive statistics) acquired through a dedicated training phase. This strategy has two main drawbacks: (i) it reduces the communication rate since no data can be transmitted during the training phase and (ii)  it can be evaded by smart or multi-antenna jammers that are quiet during the training phase or that dynamically change their subspace through time-varying beamforming. 
To address these drawbacks, we propose \emph{joint jammer mitigation and data detection (JMD)}, a novel paradigm for MIMO jammer mitigation. The core idea is to estimate and remove the jammer interference subspace jointly with detecting the transmit data over multiple time slots. Doing so removes the need for a dedicated rate-reducing training period while enabling the mitigation of smart and dynamic multi-antenna jammers. 
We instantiate our paradigm with SANDMAN, a simple and practical algorithm for multi-user MIMO uplink JMD.
Extensive simulations demonstrate the efficacy of JMD, and of SANDMAN in particular, for jammer mitigation. 

\end{abstract}

\section{Introduction}

In a world that has become fundamentally reliant on wireless communications, 
averting the threat of jamming attacks has turned into a problem of critical importance~\cite{topgun, threatvectors2021cisa, spacethreat2022, economist2021satellite}.
An attractive solution is offered by multi-antenna (MIMO) processing, which enables 
the mitigation of jammers through spatial filtering \cite{pirayesh2022jamming}.
Traditionally, a training period is used to estimate the jammer subspace, 
and the jammer's interference is removed by projecting subsequent receive signals onto the orthogonal 
complement of that subspace~\cite{shen14a, marti2021snips, yan2016jamming, do18a, akhlaghpasand20a,nguyen2021anti, leost2012interference}.
This strategy has two major disadvantages: 
First, estimating the jammer subspace during a training period reduces the achievable data rate, since no data
can be transmitted in the meantime. 
Second, estimating the jammer subspace during a training period is 
ineffective against smart jammers that transmit only at specific 
instances to evade estimation \cite{marti2022somaed} or against jammers that
change their subspace dynamically through time-varying multi-antenna transmit beamforming \cite{hoang2021suppression}.
To overcome these limitations, we propose a novel paradigm for jammer mitigation through MIMO processing which we call \emph{joint jammer mitigation and data detection (JMD)}.

\subsection{State of the art}
The fundamental challenge of jammer mitigation through spatial filtering is that
it requires information about the jammer, 
such as the subspace spanned by the jammer's channel \cite{yan2016jamming, shen14a, marti2021snips, do18a} 
or the covariance matrix of the jammer's interference \cite{marti2021snips, zeng2017enabling, marti2021hybrid}.
Existing results often assume that the jammer transmits permanently (and with static signature).
This would enable the receiver to estimate the required quantities during a dedicated training period in which the legitimate transmitters do not transmit 
\cite{shen14a, marti2021snips} or in which they transmit predetermined symbols that carry no information  
\cite{yan2016jamming,do18a, akhlaghpasand20a,nguyen2021anti,leost2012interference}. 
By relying on such a  static jammer assumption,
the receiver can filter the jammer in the subsequent communication period until the wireless channel changes and the process 
is started anew.
A smart jammer, however, can circumvent such mitigation methods by deliberately violating 
their assumptions: It can pause jamming for the duration of the dedicated training period, 
so that the receiver learns nothing meaningful~\cite{marti2022somaed}.
Or, if the jammer has multiple antennas, it can use beamforming to dynamically change its subspace
(as well as the interference covariance matrix at the receiver), so that, after the training period, 
the receiver's filter will no longer match the jammer's transmit characteristics 
\cite{hoang2021suppression}.\footnote{Another hard-to-mitigate threat is posed by multiple 
single-antenna distributed jammers, 
which also cause high-rank interference \cite{gulgun2020massive, vinogradova16a}.}
To mitigate smart jammers, methods have been suggested that attempt to fool the jammer
into transmitting during the training period by distributing and randomizing the timing of the training 
period \cite{shen14a, yan2016jamming}. However, such methods may not work against jammers that jam only 
intermittently at random time instants.
Similarly, methods have been proposed to mitigate dynamic multi-antenna jammers by recurrently 
estimating their instantaneous subspace~\cite{hoang2021suppression}, but this may be effective only against jammers that change 
their subspace in a sufficiently slow manner. 
Furthermore, all training-period based mitigation methods are subject to an inherent trade-off between the time that 
they dedicate to the attempt of estimating the required jammer characteristics 
and the time that remains for payload data transmission. 

In light of these considerations, there is a clear need for a more principled 
approach to MIMO-based jammer mitigation. 
We have recently proposed MAED \cite{marti2022somaed}, which, in hindsight, can be viewed as a special 
case of JMD. 
MAED unifies not only jammer mitigation and data detection,
but also channel estimation---at the cost of high computational complexity.

\subsection{Contributions}

We propose \emph{joint jammer mitigation and data detection (JMD)}, a novel paradigm for jammer mitigation. 
The core~idea is to estimate and remove the subspace of the jammer interference \emph{jointly}
with detecting the data of an entire transmission frame (or coherence interval). 
JMD removes the need for a dedicated training period and enables higher data rates.
Beyond that, considering an entire transmission 
period at once enables JMD to deal with smart jammers that try to evade mitigation (i)~by jamming		 only at specific instances  or (ii) by dynamically changing their subspace through multi-antenna~beamforming.

As in \cite{marti2022somaed}, we exploit the fact that a jammer cannot leave its subspace within a 
coherence interval (this holds also for multi-antenna jammers with time-varying beamforming).
Going beyond our work~in~\cite{marti2022somaed}, we show that this fact can 
be leveraged to mitigate contamination of the channel estimate even when the channel is 
estimated separately from jammer mitigation and data detection. 
This new insight opens the door to a wide range of efficient signal~processing algorithms for jammer mitigation. 
We capitalize on it by proposing SANDMAN (short for SimultANeous Detection and MitigAtioN),
an algorithm that offers all of the jammer mitigation capabilities of the MAED method from \cite{marti2022somaed}, 
but at reduced computational complexity. 
Furthermore, SANDMAN can mitigate distributed and multi-antenna jammers, while MAED only mitigates 
single-antenna jammers.
This paper also goes beyond~\cite{marti2022somaed}~by analyzing the rate improvements offered by 
the lack of a jammer estimation phase. 
Extensive simulations show the efficacy of JMD, and of SANDMAN in particular.
An extended journal version with theoretical foundations, more detailed descriptions, 
and further empirical evaluation is in preparation.

\section{Joint jammer mitigation and data detection}
\subsection{System model}
We focus on mitigating jamming attacks in the (massive) MU-MIMO uplink,
although our methods are easily translatable to other MIMO contexts. 
Consider a jamming attack such that the receive signal at the BS is given by
\begin{align}
	\bmy_k = \bH \bms_k + \Hj \bsj_k + \bmn_k.  \label{eq:model}
\end{align}
Here, $\bmy_k\in\opC^B$ is the BS receive vector at time $k$,
$\bH \in \opC^{B\times U}$ is the UE channel matrix (we assume block fading with~coherence
time $K=D+U$), 
$\bms_k\in\setS^U$ contains the time-$k$ transmit symbols of $U$ single-antenna UEs
with constellation $\setS$, 
$\bJ \in\opC^{B\times I}$ is the channel matrix of an \mbox{$I$-antenna} jammer,\footnote{The
model in \eqref{eq:model} can also represent distributed 
single- or multi-antenna jammers with a total of $I$ antennas. We consider this case in \fref{sec:multi}.} 
$\bmw_k\in\opC^{I}$ is the time-$k$ jammer transmit vector, and \mbox{$\bmn_k\sim\setC\setN(\mathbf{0},N_0\bI_B)$}
is circularly-symmetric complex 
Gaussian noise with per-entry variance $N_0$.
In this paper, we consider~$\setS$ to be QPSK (though larger constellations would also work)
scaled to unit symbol power so that $\Ex{}{\bms_k\herm{\bms_k}}=\bI_U$.
The~jammer is a dynamic multi-antenna jammer, meaning that it can dynamically change
its jamming activity.
Specifically, the jammer can transmit
\begin{align} 
	\bsj_k = \bA_k \tilde\bsj_k, \label{eq:jammer_beamforming}
\end{align}
where, without loss of generality, the covariance of $\tilde\bsj_k \!\in\! \opC^I$ is~$\bI_I$ for all $k$, 
and $\bA_k\in\opC^{I\times I}$ is a beamforming matrix~that~can change arbitrarily over time, i.e., $\bA_k$ depends on $k$.
In particular, $\bA_k$ can sometimes be the all-zero matrix (=\,no jamming), some of its rows can be zero (=\,the jammer uses only a subset of its antennas), or it can be rank-deficient in some other~way. 

\subsection{Joint jammer mitigation and data detection}
Existing methods typically null a jammer by 
projecting the receive signal onto the orthogonal complement of
the jammer subspace through $\bP\bmy_k$, where\footnote{
In this paper, $^\dagger$ denotes the pseudo-inverse, $\text{col}(\bM)$ 
is the column space of $\bM$, and $^\orth$ denotes the orthogonal complement.} $\bP = \bI_B - \Hj\pinv\Hj$
is the orthogonal projection onto $\text{col}(\Hj)^\orth$, 
which has the property that $\bP\bJ=\mathbf{0}$.
After the projection, the data can be detected using the virtual channel matrix $\bH_\bP = \bP\bH$, 
since
\begin{align}
	\bP\bmy_k &= \bP\bH\bms_k + \bP\Hj\bA_k \tilde \bsj_k + \bP \bmn_k 
	= \bP\bH\bms_k + \bP \bmn_k \\
	&\triangleq \bH_\bP \bms_k + \bmn_{\bP,k}.
\end{align}
Note that this works regardless of which vectors $\tilde\bsj_k$ and which matrices
$\bA_k$ the jammer uses, since $\text{col}(\bJ)\supseteq\text{col}(\bJ\bA_k)$ and so
$\text{col}(\bJ)^\orth \subseteq \text{col}(\bJ\bA_k)^\orth$ for any $\bA_k$. 
The problem, however, is how to reliably estimate $\bJ$---or $\text{col}(\bJ)$---when the jammer changes $\bA_k$ dynamically such that $\text{col}(\bJ\bA_k)$ depends on $k$. 

The central idea of JMD is to consider jammer subspace estimation, jammer mitigation, 
and data detection over an entire transmission frame (or coherence interval) simultaneously: The jammer subspace is identified with the subspace that is not
explainable in terms of UE transmit signals, which are estimated iteratively while projecting the receive signals onto the orthogonal
complement of the current estimate of the jammer subspace.
Mathematically, this can be framed as~solving
\begin{align}
\textstyle
	\min_{\tilde\bS_D, \tilde\bP}\, \| \tilde\bP \bY_D - \tilde\bP \bH \tilde\bS_D \|_F^2, \label{eq:jmd_problem}
\end{align}
where $\bY_D\!=\! [\bmy_1, \dots, \bmy_D]$ is the data receive matrix over an entire coherence interval, 
$\tilde\bS_D = [\tilde\bms_1, \dots, \tilde\bms_D]\!\in\! \setS^{U\times D}$ is the data matrix estimate, 
and $\tilde\bP = \bI_B - \tilde\bJ\pinv{\tilde\bJ}$ is the projection onto the orthogonal
complement of the estimated jammer subspace $\text{col}(\tilde\bJ)$, with $\tilde\bJ\in\opC^{B\times I}$.
The range over which we optimize $\tilde\bP$ is the Grassmanian manifold $\mathscr{G}_{B-I}(\opC^B)$, i.e.,~the 
set of orthog-onal projections onto $(B\!-\!I)$-dimensional subspaces~of~$\opC^B$.

In practice, however, the UE channel matrix $\bH$ is \textit{a priori} unknown and has to be
estimated with pilots. In the presence of jamming, the obtained estimate can be contaminated. 
Thankfully, however, the contamination of the pilot receive signal will be restricted to the subspace $\text{col}(\bJ)$. 
In~\cite{marti2022somaed}, it was thus proposed to jointly solve the problems of channel estimation, jammer
subspace estimation and mitigation, and data detection by solving
an optimization problem which depends on $\tilde\bS_D, \tilde\bP,$ and $\tilde\bH_{\tilde\bP}$.
This approach is highly effective, but unfortunately, even approximately solving the proposed optimization problem 
is computationally demanding.  
For instance, the algorithm proposed in \cite{marti2022somaed} requires the inversion of a $U\times U$ matrix 
for gradient calculation in every iteration.

As it turns out, and this is a key insight of our paper, it is not necessary to
estimate $\bH$ jointly with the jammer subspace and the data symbols: One can also estimate the channel separately, 
leading to algorithms with significantly lower complexity. 
Consider least square (LS) channel estimation with unitary pilots $\bS_T\in\opC^{U\times U}$ and receive matrix~$\bY_T$:
\vspace{-1mm}
\begin{align}
	\bY_T &= \bH \bS_T + \bJ\bW + \bN \\
	\hat\bH &= \bY_T \herm{\bS_T} = \bH + \bJ \bW\herm{\bS_T} + \bN\herm{\bS_T}. \label{eq:pilot_contamination}
\end{align}
Note how after despreading, the jammer contamination of the channel estimate
is still restricted to $\text{col}(\bJ)$.\footnote{This would also be the case for other linear channel estimates 
\mbox{$\hat\bH=f(\bY_D)$}, such as the LMMSE channel estimate.} If we therefore simply plug this estimate 
$\hat\bH$ into \eqref{eq:jmd_problem} as follows
\begin{align}
\textstyle
	\min_{\tilde\bS_D, \tilde\bP}\, \| \tilde\bP \bY_D - \tilde\bP \hat\bH \tilde\bS_D \|_F^2, \label{eq:jmd_problem2}
\end{align}
then the ``true'' projection $\bP = \bI_B - \Hj\pinv\Hj$ also removes the jammer contamination of the channel estimate.
This means that we can leverage the concept of joint jammer subspace estimation, jammer mitigation, and data detection, 
without having to also jointly estimate $\bH$, and without having to worry about jammer contamination of the channel estimate. 
In summary, the JMD paradigm can be formulated as follows:

\begin{mdframed}[style=csstyle,frametitle={Joint Jammer Mitigation and Data Detection (JMD)}]
\vspace{-1mm}
Given the received data matrix $\bY_D\in\opC^{B\times D}$ of an entire coherence interval 
and a linear channel estimate \mbox{$\hat\bH=f(\bY_D)$} from that same coherence interval, 
solve
\begin{align}
	\min_{\substack{\tilde\bS_D \,\in\, \setS^{U\times D},\\ \tilde\bP \,\in\, \mathscr{G}_{B-I}(\opC^B)}} 
	\big\| \tilde\bP (\bY_D - \hat\bH \tilde\bS_D) \big\|_F^2. \label{eq:jmd}
\end{align}
\end{mdframed}
\vspace{0mm}

\section{The SANDMAN algorithm}
Solving \eqref{eq:jmd} exactly is difficult, so we solve it approximately.
A key difficulty is that, due to the discreteness of $\setS$, the problem in \eqref{eq:jmd} is
NP-hard even when fixing $\tilde\bP$ and solving only for~$\tilde\bS_D$~\cite{grotschel2012geometric}.
We thus relax the constraint set $\setS$ to its convex hull $\setC\triangleq \textit{conv}(\setS)$. 
To promote symbol estimates at, or near, the corner points of $\setC$ (i.e.,the constellation points~$\setS$), 
we add a concave regularizer $-\|\tilde\bS_D\|_F^2$ weighted by $\alpha>0$ to the objective \cite{shah2016biconvex}.
We colloquially refer to the resulting constraint and regularizer as a \emph{box prior}.
The modified problem is thus
\begin{align}
	\min_{\substack{\tilde\bS_D \,\in\, \setC^{U\times D},\\ \tilde\bP \,\in\, \mathscr{G}_{B-I}(\opC^B)}} 
	\big\| \tilde\bP (\bY_D - \hat\bH \tilde\bS_D) \big\|_F^2 - \alpha\|\tilde\bS_D\|_F^2. \label{eq:sandman_problem}	
\end{align}
This problem is still non-convex, mainly due to the non-convex constraint set $\mathscr{G}_{B-I}(\opC^B)$ of $\tilde\bP$. 
However, we have the following theorem, the proof of which is omitted due to lack of space.

\begin{thm} \label{thm:1}
	When $\tilde\bP$ is fixed and $\alpha\leq \lambda_{\min}((\tilde\bP\hat\bH)^{\textnormal{H}}\tilde\bP\hat\bH)$, 
	then the objective in \eqref{eq:sandman_problem} is convex in $\tilde\bS_D$.
	Vice versa, when $\tilde\bS_D$ is fixed, then the objective in \eqref{eq:sandman_problem} is minimized with 
	respect to $\tilde\bP$ by $\bI_B - \bU_I\herm{\bU_I}$, where $\bU_I\in\opC^{B\times I}$ consists of the $I$ 
	dominant left-singular vectors of $\bY_D - \hat\bH \tilde\bS_D$.
\end{thm}
\vspace{-1mm}
This theorem suggests to use an alternating minimization strategy, as solving \eqref{eq:sandman_problem}
for either $\tilde\bS_D$ or $\tilde\bP$ is straightforward while the other quantity is fixed. 

\subsubsection{Solving for $\tilde\bS_D$}
To solve the problem in \eqref{eq:sandman_problem} for $\tilde\bS_D$, we use forward-backward splitting (FBS) \cite{goldstein16a}. 
FBS is a method for iteratively solving convex optimization problems of 
the~form
\begin{align}
	\min_{\tilde\bms}~ f(\tilde\bms) + g(\tilde\bms), \label{eq:fbs_problem}
\end{align}
where $f$ is convex and differentiable, and $g$ is convex but need not be differentiable, smooth, or bounded. 
FBS solves the problem in \eqref{eq:fbs_problem} by iteratively computing 
\begin{align}
	\tilde\bms^{(t+1)} = \proxg\big(\tilde\bms^{(t)} - \tau^{(t)}\nabla f(\tilde\bms^{(t)}); \tau^{(t)}\big), \label{eq:fbs_iteration}
\end{align}
where $\tau^{(t)}$ is the stepsize at iteration $t$, $\nabla f$ is the gradient of~$f$,  
and $\proxg$ is the proximal operator of $g$, defined as \cite{parikh13a}
\begin{align}
\proxg(\tilde\bms; \tau) = \arg\!\min_{\tilde\bmx~~~~} \tau g(\tilde\bmx) + \frac12 \|\tilde\bms - \tilde\bmx\|_2^2. 
\end{align}
FBS solves convex problems exactly (for a sufficient number of iterations with suitable stepsizes~$\tau^{(t)}$), 
but it is also effective for approximately solving non-convex problems \cite{goldstein16a}.
To solve the problem in \eqref{eq:sandman_problem}, we define the functions $f$ and~$g$ as
\begin{align}
	f(\tilde\bS_D) &= \big\| \tilde\bP (\bY_D - \hat\bH \tilde\bS_D) \big\|_F^2, \\
	g(\tilde\bS_D) &= - \alpha\big\|\tilde\bS_D\big\|_F^2 + \chi_\setC(\tilde\bS_D), 
\end{align}
where $\chi_\setC$ acts entrywise on $\tilde\bS_D$ as the indicator function of $\setC$,
\begin{align}
	\chi_\setC(\tilde s) = \begin{cases}
		0 &: \tilde s \in \setC \\
		\infty &: \tilde s \notin \setC.
	\end{cases}
\end{align}
The gradient of $f$ in $\tilde\bS_D$ is given as 
\begin{align}
	\nabla f(\tilde\bS) = -2\,\herm{\hat\bH}\tilde\bP(\bY_D - \hat\bH \tilde\bS).
\end{align}
The proximal operator of $g$ acts entrywise on $\tilde\bS_D$ and is given~as
$\proxg(\tilde s; \tau) = \text{clip}(\tilde s/(1-\tau\alpha); \sqrt{\sfrac{1}{2}})$ when $\alpha\tau<1$ 
(where $\text{clip}(z;a)$ clips the real and imaginary part of $z\in\opC$ to the~interval $[-a,a]$),
and otherwise as 
$\arg\!\min_{\tilde x \in \{\pm\sqrt{\frac{1}{2}} \pm i\sqrt{\frac{1}{2}} \} } |\tilde s - \tilde x|^2$.

\subsubsection{Solving for $\tilde\bP$}
According to \fref{thm:1}, we can solve for~$\tilde\bP$ (for fixed $\tilde\bS_D$) 
by calculating the $I$ dominant left-singular vectors $\bU_I$ of $\bY_D - \hat\bH \tilde\bS_D$.
Instead of performing an exact but computationally expensive singular value decomposition
of this matrix, we approximate $\bU_I$ with the power method from~\cite{liberty2013svd}, 
where we perform a single power iteration per~dimension. 
\vspace{2mm}

The SANDMAN algorithm alternates between descent steps in $\tilde\bS_D$ and approximate computations of $\tilde\bP$ 
for a total number of $t_{\max}$ iterations. We choose $\alpha=2.5$, and the stepsizes $\tau^{(t)}$
are selected using the Barzilai-Borwein criterion detailed in \cite{zhou2006gradient}.
SANDMAN is summarized in \fref{alg:sandman} and has a complexity of
$O(t_{\max}UDB)$, i.e., its complexity is linear in $U$, $D$, and $B$.
 
\floatstyle{spaceruled}
\restylefloat{algorithm}
\begin{algorithm}[tp]
  \caption{SANDMAN}
  \label{alg:sandman}
  \begin{algorithmic}[1]
	\setstretch{1.0}
	\vspace{1mm}
    \Function{SANDMAN}{$\bY_D, \bY_T, \bS_T, I, t_{\max}$}
    \State  $\hat\bH = \bY_T \herm{\bS_T}$ \hfill \textcolor{gray}{// LS channel estimate}
    \State $\tilde\bS^{(0)} = \mathbf{0}_{U\times D}$
    \For{$t=0$ {\bfseries to} $t_{\max}-1$}
		\State $\tilde\bE^{(t)} = [\bY_T,\bY_D] - \hat\bH [\bS_T,\tilde\bS^{(t)}]$
		\State $\tilde\bJ^{(t)} = \textsc{approxSVD}(\tilde\bE^{(t)}, I)$ 
		\hfill\textcolor{gray}{// cf. \cite{liberty2013svd}} 
		\State $\tilde\bP^{(t)} = \bI_B - \tilde\bJ^{(t)}\pinv{(\tilde\bJ^{(t)})}$
		\State $\nabla f(\tilde\bS^{(t)}) = -2\,\herm{\hat\bH}\tilde\bP^{(t)}(\bY_D - \hat\bH \tilde\bS^{(t)})$
		\State $\tilde\bS^{(t+1)} = \proxg\big(\tilde\bS^{(t)} - \tau^{(t)}\nabla f(\tilde\bS^{(t)}); \tau^{(t)}\big)$
    \EndFor
    \State \textbf{output:} $\tilde\bS^{(t_{\max})}$    
    \EndFunction    
  \end{algorithmic}
\end{algorithm}

\section{Evaluation}

\subsection{Simulation setup}
We evaluate SANDMAN through simulations.  
We simulate a \mbox{MU-MIMO} system with $B=32$ BS antennas and $U=16$ single-antenna UEs
at a carrier frequency of 2\,GHz 
using the 3GPP 38.901 urban macrocellular (UMa) channel model~\cite{3gpp22a}. 
The channel vectors are generated with QuaDRiGa \cite{jaeckel2014quadriga}.
The BS antennas are arranged as a uniform linear array (ULA) and spaced at half wavelength.
The UEs are uniformly distributed at distances between $10$\,m and $250$\,m in a $120^\circ$ angular sector 
in front of the BS, and with a minimum angular separation of~$1^\circ$ between any two UEs. 
All antennas are omnidirectional.
We assume $\pm3$\,dB per-UE power control.
Furthermore, we assume a coherence time of $K=100$ channel uses. 
The specific jammer model varies between the different experiments. In general, we consider 
$J\geq1$ jammers placed randomly in the same area as the UEs, with a minimum angular separation 
of $1^\circ$ between any two jammers as well as between any jammer and any UE. 
Every jammer is equipped with $I/J\geq1$ antennas arranged as a ULA with half-wavelength 
spacing that is frontally facing in the direction of~the~BS. 

We consider QPSK transmission. 
The pilots are selected as rows of a $U\times U$ Hadamard matrix (normalized to unit symbol energy). 
We define the average signal-to-noise ratio (SNR) as
\begin{align}\textstyle
\textit{SNR} \define \frac{\Ex{\bS}{\|\bH\bS\|_F^2}}{\Ex{\bN}{\|\bN\|_F^2}}.
\end{align}
Furthermore, we characterize the strength of the jammer interference relative to the 
strength of the average UE via
\begin{align}\textstyle
	\rho \define \frac{\|\bJ\bW\|_F^2}{\frac1U\Ex{\bS}{\|\bH\bS\|_F^2}},
\end{align}
where we deterministically scale $\bJ\bW$ to a~pre-specified $\rho$. 
As performance metrics, we consider uncoded bit error rate (BER) and 
a metric that we call the modulation error ratio (MER) between the data symbols $\bS_D$ 
and their estimate~$\hat\bS_D$, 
\begin{align}
	\textit{MER}\triangleq \mathbb{E}\big[\|\hat\bS_D - \bS_D\|_F\big]\big/\mathbb{E}\big[\|\bS_D\|_F\big].
\end{align}
We use the MER as a surrogate for error vector magnitude (EVM), which 
the 3GPP 5G NR technical specification requires to be below 17.5\% \cite[Tbl. 6.5.2.2-1]{3gpp21a} for QPSK transmission.

\subsection{Higher data rates against simple jammers}\label{sec:rate}
\begin{figure}[tp]
\centering
\includegraphics[height=3.6cm]{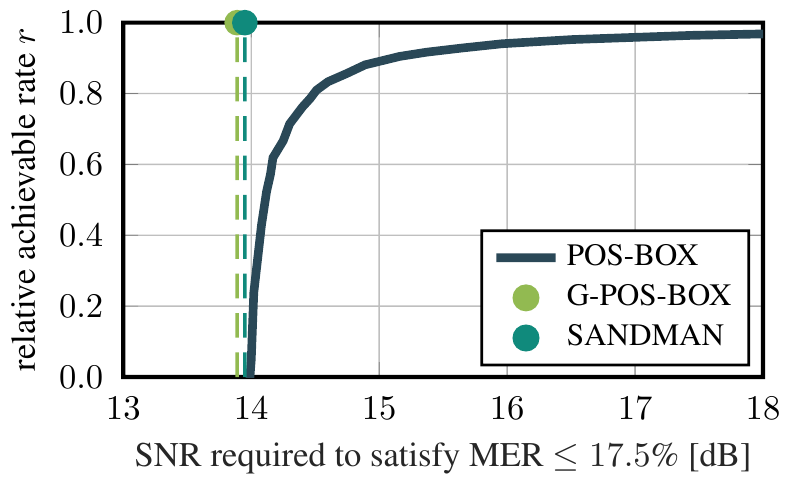}
\caption{Trade-off between the relative achievable rate $r$ and the smallest SNR for which the different 
receivers satisfy the criterion $\text{MER}\leq17.5\%$ 
when mitigating a single-antenna barrage jammer \tinygraycircled{1}. 
}
\label{fig:rate_reduction}
\end{figure}

The first advantage of joint jammer mitigation and data detection is increased achievable rates, 
because no channel uses need to be reserved for estimating the jammer's subspace.

\vspace{2mm}
\subsubsection*{\textbf{Jammer model}}
The rate advantage afforded by the absence of a jammer estimation phase is shown on the following model:
\subsubsection*{\graycircled{\textnormal{1}} Strong single-antenna barrage jammer} 
In this model, a single jammer with a single antenna transmits i.i.d. circularly-symmetric complex
Gaussian noise over the entire coherence interval, with a relative receive strength that exceeds 
the receive strength of the average UE by $\rho=30$\,dB.

\vspace{2mm}
\subsubsection*{\textbf{Baselines}} 
The following receivers are used as baselines:

\subsubsection*{G-POS-BOX}
This receiver serves as a performance upper bound. It 
works analogous to SANDMAN but is furnished with ground-truth knowledge of the jammer channel $\bJ$, 
and fixes $\tilde\bP^{(t)}$ in line 7 of \fref{alg:sandman} to the optimal projector $\bP = \bI_B - \Hj\pinv\Hj$.

\subsubsection*{POS-BOX}
This receiver uses a protocol in which, at the beginning of every coherence interval, the UEs do not 
transmit for $L$ channel uses (at cost of a reduced $D=K-U-L$). The receive matrix $\bY_J\!\in\!\opC^{B\times L}$ from this 
period is used to estimate the jammer subspace as the strongest~$I$ left singular vectors of $\bY_J$. 
The jammer is then mitigated by projecting subsequent receive signals onto the orthogonal complement of the 
estimated subspace, and by performing LS channel estimation and FBS-based data detection with 
a box prior (analogous to SANDMAN)
in that projected space. 

All algorithms run for $t_{\max}=30$ iterations.

\vspace{2mm}
\subsubsection*{\textbf{Results}}
\fref{fig:rate_reduction} depicts the results. 
To assess the rate~advantage afforded by the absence of a jammer estimation phase, we 
consider the tradeoff between the SNR threshold for which a receiver
satisfies the criterion $\text{MER}\leq 17.5\%$ and the ratio
\begin{align}
	r = \frac{K-U-L}{K-U}
\end{align}
that is the number of samples $K-U-L$ available per coherence interval for data transmission, 
normalized by the maximum number of samples $K-U$ available when using orthogonal pilots and no jammer estimation phase.
Clearly,~$r$ translates directly to achievable data rate (in bits/s).
Both SANDMAN and G-POS-BOX do not have a jammer estimation phase ($L=0$) and hence have $r=1$. 
In contrast, the performance of \mbox{POS-BOX} improves as $L$ is increased
to obtain a better estimate of the jammer subspace, at the expense of~$r$.
The results show that SANDMAN, which utilizes the receive data of the entire coherence interval to estimate the 
jammer subspace, achieves virtually the same performance as G-POS-BOX, with the required SNR differing only by about $0.1$\,dB. 
POS-BOX, in contrast, can use only a subset of the receive samples to estimate the jammer subspace, and so performs much worse even when sacrificing a large part of the coherence time for jammer estimation.
To approach the performance of SANDMAN up to $0.5$\,dB, G-POS-BOX has to accept a $20$\% rate reduction. 
G-POS-BOX does not even outperform SANDMAN when using virtually the \emph{entire} coherence interval 
for jammer estimation~(at the expense of a vanishing rate $r$).
This shows that SANDMAN can leverage the same receive data \emph{twice}---for jammer 
esti-mation~\emph{and} for data detection---without impairing performance.

\begin{figure*}[tp]
\centering
\subfigure[barrage jammer \tinygraycircled{1} \hspace{-9mm}]{
\includegraphics[height=3.8cm]{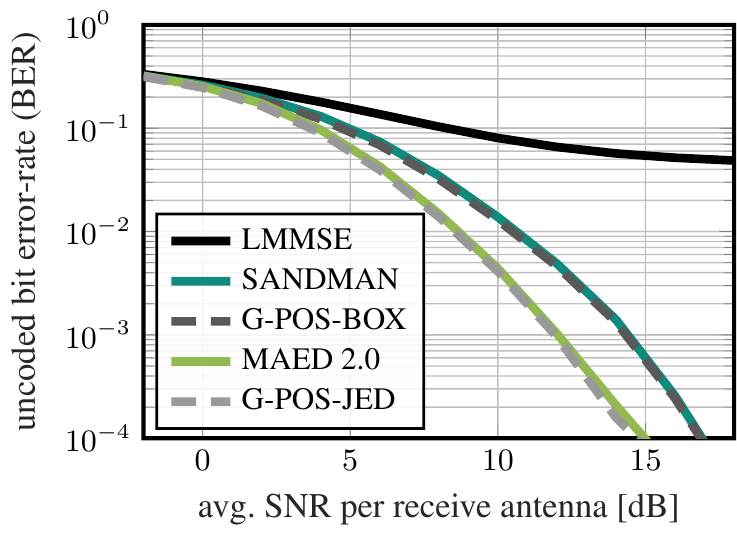}
}\qquad
\subfigure[data jammer \tinygraycircled{2} \hspace{-10mm}]{
\includegraphics[height=3.8cm]{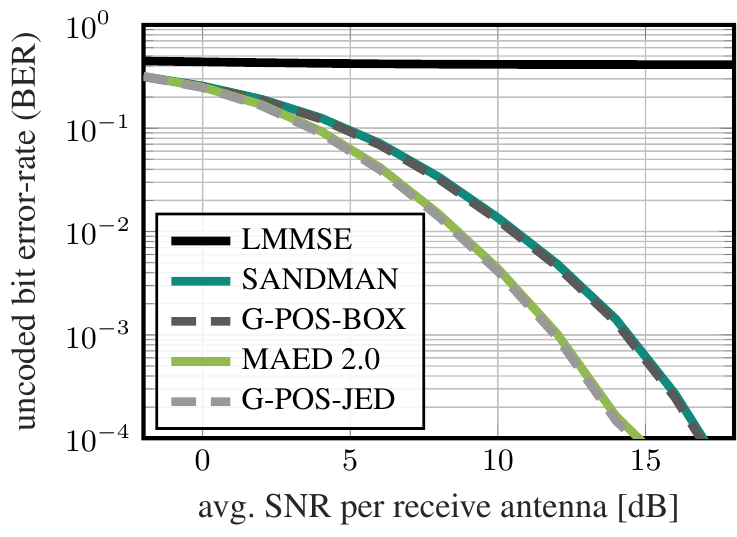}
}\qquad
\subfigure[pilot jammer \tinygraycircled{3} \hspace{-10mm}]{
\includegraphics[height=3.8cm]{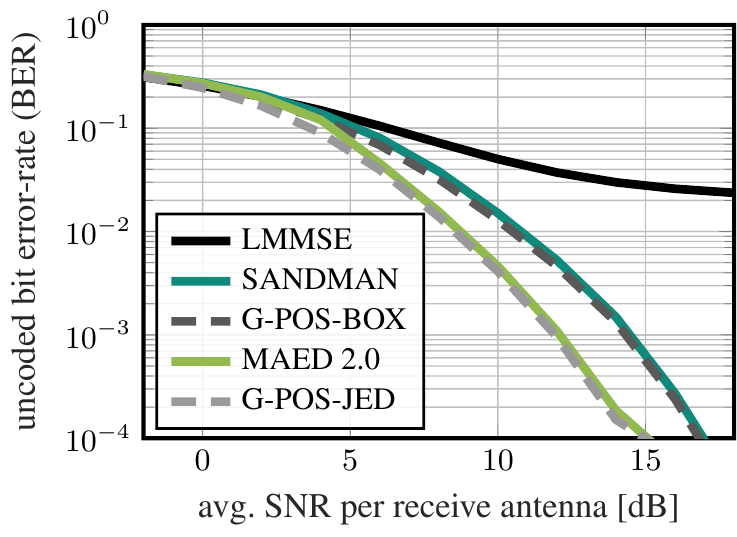}
}
\caption{Uncoded bit error-rate (BER) vs. SNR performance of different receivers when mitigating different kinds of smart
single-antenna jammers.}
\label{fig:smart}
\vspace{-2mm}
\end{figure*}

\begin{figure*}[tp]
\centering
\subfigure[distributed barrage jammers \tinygraycircled{4} \hspace{-8mm}]{
\includegraphics[height=3.8cm]{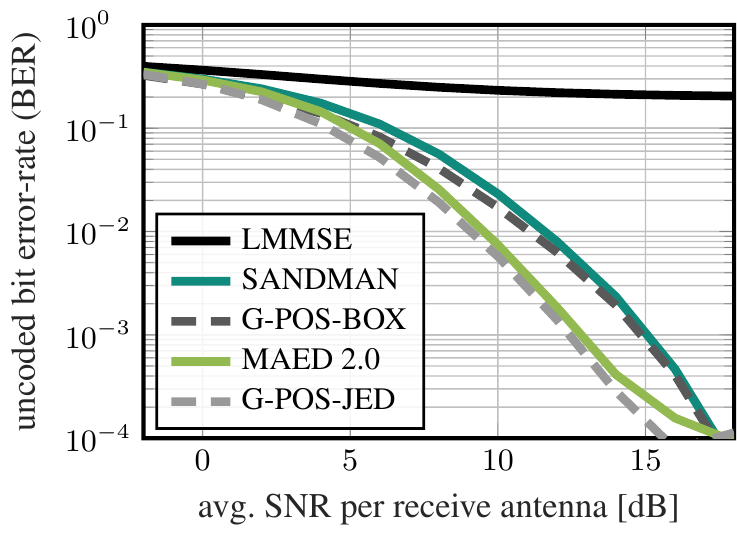}
}\qquad
\subfigure[jump-varying beamformer  \tinygraycircled{5} \hspace{-9mm}]{
\includegraphics[height=3.8cm]{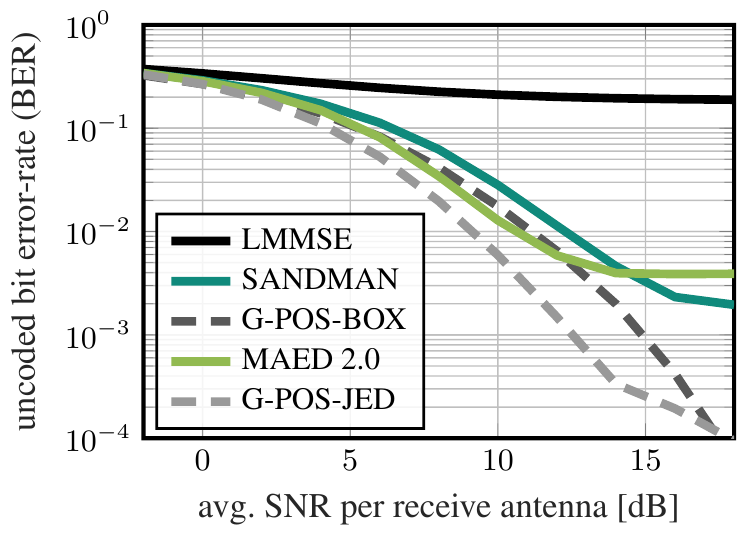}
}\qquad
\subfigure[continuously-varying beamformer \tinygraycircled{6} \hspace{-8mm}]{
\includegraphics[height=3.8cm]{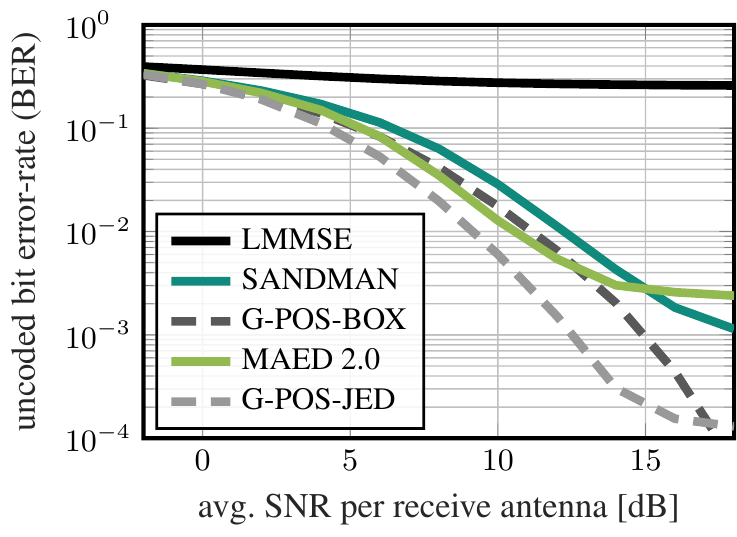}
}
\caption{Uncoded bit error-rate (BER) vs. SNR performance of different receivers when mitigating 
distributed jammers or dynamic multi-antenna jammers.}
\label{fig:multi}
\vspace{-2mm}
\end{figure*}

\subsection{Mitigating smart single-antenna jammers} \label{sec:smart}
In this experiment, we analyze SANDMAN's ability to mitigate smart jammers that 
would suspending jamming during a potential training phase, and that target only specific parts 
(such as the pilot phase or the data phase) of the transmission.\nobreak
\vspace{2mm}
\subsubsection*{\textbf{Jammer model}}
Besides a strong barrage jammer (\graycircled{1}), we consider the following types of smart (time-variant) jammers:
\subsubsection*{\graycircled{2} Strong single-antenna data jammer} A single-antenna jammer
that does not jam the pilot phase but transmits i.i.d. complex Gaussian noise during the data phase with \mbox{$\rho=30$\,dB.}
\subsubsection*{\graycircled{3} Strong single-antenna pilot jammer} A single-antenna jammer
that does not jam the data phase but transmits i.i.d. complex Gaussian noise during the pilot phase with \mbox{$\rho=30$\,dB.}

\vspace{2mm}
\subsubsection*{\textbf{Baselines}} 
Besides G-POS-BOX (cf.~\fref{sec:rate}), the following baselines are considered:\footnote{Training-period 
based mitigation methods are incapable of mitigating smart jammers, see \cite{marti2022somaed}, 
which is why our comparison omits POS-BOX.}
\subsubsection*{LMMSE} This receiver does not mitigate the jammer and so provides a lower bound on performance. 
It performs LS channel estimation and (jammer-oblivious) LMMSE data detection. 
\subsubsection*{MAED 2.0} This is an extension of MAED \cite{marti2022somaed} for single- 
\emph{and} multi-antenna jammers. 
In contrast to SANDMAN, it uses joint channel estimation and data detection (JED), which leads to better
performance at increased computational cost \cite{vikalo2006efficient}.\footnote{A 
full description of MAED 2.0 will be provided in an extended journal version of this paper which is in preparation.}
\subsubsection*{G-POS-JED} A performance upper bound for MAED 2.0.~It works analogous to MAED 2.0 (using JED) but
uses ground-truth knowledge of $\bJ$ for the optimal projector $\bP = \bI_B - \Hj\pinv\Hj$.

All iterative algorithms run for $t_{\max}=30$ iterations.

\vspace{2mm}
\subsubsection*{\textbf{Results}}
\fref{fig:smart} depicts the results. As expected, the non-mitigating LMMSE receiver always has by far the worst performance, 
though it does not suffer equally under all jammers. G-POS-BOX and the G-POS-JED use ground-truth knowledge
to null the jammer perfectly in all cases, so that their performance does not depend on the jammer transmit signal. 
Consequently, their BER curves are identical for all types of single-antenna jammers (\graycircled{1}\,-\,\graycircled{3}), 
with G-POS-JED slightly outperforming G-POS-BOX due to the advantage of JED over separate channel estimation and data detection. 
The results show that SANDMAN and MAED 2.0 both achieve virtually the same performance as their respective 
performance upper bounds (which rely on ground-truth jammer knowledge) for all simulated jammer types.
This implies that SANDMAN and MAED 2.0 are able to estimate the jammer subspace  essentially perfectly, regardless of 
when the jammer is active.

\subsection{Mitigating distributed jammers and multi-antenna jammers} \label{sec:multi}
\subsubsection*{\textbf{Jammer model}} We consider both distributed 
and multi-antenna jammers. The difference is that distributed jammers
cannot form beams while multi-antenna jammers can use dynamic beamforming to change their subspace 
but are~colocated. 
\subsubsection*{\graycircled{4} Distributed barrage jammers} 
We consider four distributed single-antenna jammers that transmit
(independent of each other) i.i.d. Gaussian noise with $\rho=30$\,dB
(each jammer transmits at $24$\,dB more receive power than the average UE). 
\subsubsection*{\graycircled{5} Jump-varying beamforming jammer}
We consider a single four-antenna jammer that, at every instant $k$, transmits
only~on a random subset of between one and three of its antennas. This is 
achieved by selecting only a random subset of the rows~of its beamforming matrix 
$\bA_k$ (see \eqref{eq:jammer_beamforming}) to be nonzero 
(nonzero rows have i.i.d. Gaussian entries), 
and by switching to a completely new matrix $\bA_k$ at random 
instances $k_1,\dots,k_M, M=5$. The jamming vectors $\tilde\bmw_k$ have distribution 
$\{\tilde\bmw_k\}\stackrel{\text{i.i.d.}}{\sim}\setC\setN(\mathbf{0},\bI_I)$.
\subsubsection*{\graycircled{6} Continuously-varying beamforming jammer}
This (single) jammer also has four antennas. 
Only the leftmost column $\bma_{k,1}$ of its beamforming matrix $\bA_k$ (see \eqref{eq:jammer_beamforming})
is nonzero. For randomly selected instances $k_1,\dots,k_M, M=5$, with $k_m<k_{m+1}$,
the vector $\bma_{k,1}$ is fixed to randomly and independently drawn vectors $\{\bma^{(m)}\}_{m=1}^M$. 
For $k_m<k<k_{m+1}$, $\bma_{k,1}$ interpolates smoothly between $\bma^{(m)}$ and $\bma^{(m+1)}$.
The jamming vectors $\tilde\bmw_k$ have distribution 
$\{\tilde\bmw_k\}\stackrel{\text{i.i.d.}}{\sim}\setC\setN(\mathbf{0},\bI_I)$.

\vspace{2mm}
\subsubsection*{\textbf{Baselines}} We consider the same baselines as in \fref{sec:smart}.
All iterative algorithms run for $t_{\max}=50$ iterations.

\vspace{2mm}
\subsubsection*{\textbf{Results}}
\fref{fig:multi} depicts the results. The LMMSE receiver has again by far the worst performance with double-digit~BER percentages. Because they have to null four dimensions, G-POS-JED and G-POS-BOX
perform slightly worse than in \fref{sec:smart} (where they only have to project away one dimension), 
but they again null all jammers perfectly. 
In case of the distributed barrage jammers (\graycircled{4}), SANDMAN and MAED 2.0 perform again very 
close to their respective performance bounds, meaning that they mitigate the jammers almost perfectly. 
SANDMAN and MAED 2.0 are also able to successfully mitigate the dynamic multi-antenna 
jammers~\mbox{(\graycircled{5}, \graycircled{6})}: They achieve BERs significantly below $1\%$ at high SNR, 
even if the BER eventually levels  off a bit above $0.1\%$. But the jammers \graycircled{5} and \graycircled{6}
are incredibly difficult to mitigate, since the jamming subspace is changed repeatedly, with some
subspaces potentially being used only for extremely brief amounts of~time.

\section{Conclusions}
We have proposed joint jammer mitigation and data detection (JMD), a novel paradigm 
for mitigating jammers in MIMO systems that does not use a dedicated jammer training period. 
As a result, JMD is able to mitigate smart jammers regardless of (i) when they 
are active
and (ii)  how they vary their multi-antenna transmit beamforming. 
We then have proposed SANDMAN, an efficient JMD-type algorithm for jammer mitigation in the \mbox{MU-MIMO} uplink.
Simulation results for a variety of different jammers types have demonstrated the utility of the JMD paradigm, 
and of SANDMAN in particular.


\vfill

\end{document}